# Graphene-based thermopneumatic generator for on-board pressure supply of soft robots

Armin Reimers[1], Jannik Rank[1], Erik Greve[1], Morten Möller[1], Sören Kaps[1], Jörg Bahr[1], Rainer Adelung[1*], Fabian Schütt[1*]

[1]*Functional Nanomaterials, Department of Materials Science, Kiel University, Kaiserstr. 2, 24143 Kiel, Germany*

*\*Corresponding Author*

## Abstract

Various fields, including medical and human interaction robots, gain advantages from the development of bioinspired soft actuators. Many recently developed grippers are pneumatics that require external pressure supply systems, thereby limiting the autonomy of these robots. This necessitates the development of scalable and efficient on-board pressure generation systems. While conventional air compression systems are hard to miniaturize, thermopneumatic systems that joule-heat a transducer material to generate pressure present a promising alternative. However, the transducer materials of previously reported thermopneumatic systems demonstrate high heat capacities and limited surface area resulting in long response times and low operation frequencies. This study presents a thermopneumatic pressure generator using aerographene, a highly porous (>99.99 %) network of interconnected graphene microtubes, as lightweight and low heat capacity transducer material. An aerographene pressurizer module (AGPM) can pressurize a reservoir of 4.2 cm$^3$ to ~140 mbar in 50 ms. Periodic operation of the AGPM for 10 s at 0.66 Hz can further increase the pressure in the reservoir to ~360 mbar. It is demonstrated that multiple AGPMs can be operated parallelly or in series for improved performance. For example, three parallelly operated AGPMs can generate pressure pulses of ~215 mbar. Connecting AGPMs in series increases the maximum pressure achievable by the system. It is shown that three AGPMs working in series can pressurize the reservoir to ~2000 mbar in about 2.5 min. The AGPM's minimalistic design can be easily adapted to circuit boards, making the concept a promising fit for the on-board pressure supply of soft robots.

# AEROGRAPHENE TRANSDUCER MATERIAL

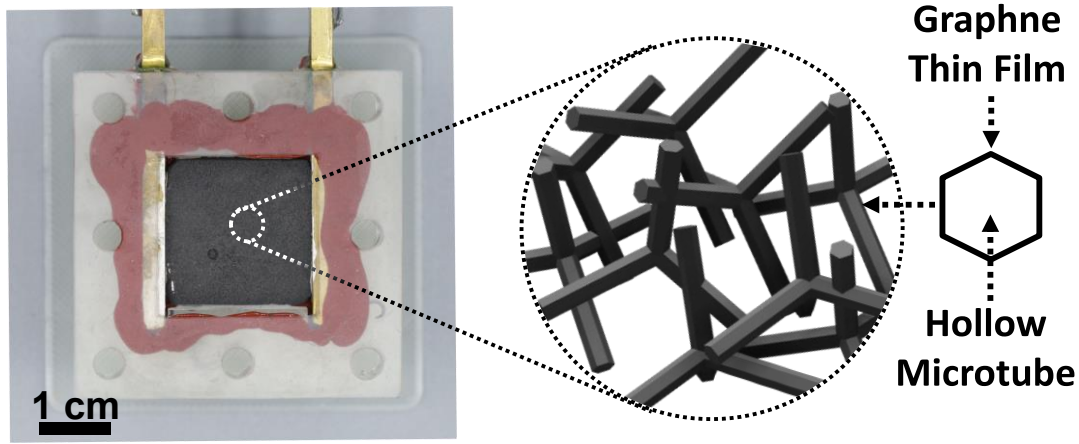

# THERMOPNEUMATIC PRESSURE GENERATOR

**TOC:** Presenting the AGPM, a thermopneumatic pressure generator using aerographene as transducer material. The aerographene enables pressure pulses with low response times and high repetition rates, making the AGPM a promising fit for the on-board pressure supply of soft robots.

# Introduction

Numerous fields stand to gain significant advantages from the development of effective actuation mechanisms employing soft materials, including medical robotics, artificial muscles, and other devices designed for human interaction.[1] Researchers are currently exploring various potential options for creating soft actuators. Much of this research is centered around soft materials capable of converting electrical energy into mechanical energy. Possible candidates are electrostrictive, dielectric or piezoelectric polymers.[2–5] However, the practicality of these actuators is constrained by their reliance on either significant electric fields or flexible electrodes.[6] Alternative approaches utilize photo- or thermoresponsive polymers that release mechanical energy when triggered by external stimuli.[7] Recent studies have presented bioinspired pneumatic grippers and actuators that can be operated by pressurized air. These devices are fabricated from structured soft materials and provide significant actuation strength when inflated with 5 mbar of pressure.[5, 8–10]

While the necessary pressurized air for the grippers can be supplied by including a gas tank in the robot design, the tank would require regular refilling, reducing the autonomy of the robot. On top, conventional air compression systems have many moving parts and still proof challenging to miniaturize, limiting the size of the robot.[11] For the full realization of autonomous mobile robots using pneumatic actuators, it is therefore necessary to develop on-board pressure generation systems.

In thermopneumatic pressure generators, a Joule-heated transducer material transfers heat to a working gas. If the volume of the working gas is confined, the resulting increase in temperature causes an increase in pressure. The relationship between pressure, volume and temperature of a gas can be sufficiently described using the ideal gas law[12]:

$$p V = nRT \qquad (1)$$

The achievable pressure and response time of a thermopneumatic system are determined by the maximum temperature of the transducer material as well as its surface area in contact with the working gas. Previously presented thermopneumatic systems use pleated or coiled metal wires to heat the working gas.[13, 14] The limited surface area of the wire restricts a rapid heat transfer to the working gas. Therefore, conventional thermopneumatic system have long response times. The high heat capacity of the wire also leads to high cooling times between Joule-heating cycles, limiting the operating frequency.[13]

This study presents a new kind of thermopneumatic pressure generator using aerographene as its transducer material. Aerographene is a hierarchical network of freestanding graphene microtubes with a porosity of >99.99 %.[15, 16] Aerographene is characterized by a low volumetric heat capacity comparable to that of air, a high electric conductivity (~50 S m$^{-1}$) and a high thermal stability.[16, 17] Furthermore, the open porous and hierarchical internal structure of the aerographene offers a large and universally accessible internal surface area which enables efficient heat transfer to the gas in and around the aerographene network structure. Previous works have demonstrated rapid Joule-heating of aerographene with heating rates exceeding 300000 K s$^{-1}$, which can be used to drive pumps as well as actuators.[16] Aerographene is therefore a promising candidate for the transducer material of thermopneumatic pressure generation. The here presented aerographene pressurizer module (AGPM) can generate high maximum pressures at low response times and high repetition rates. Multiple AGPM can be combined for enhanced pressure generation, providing an additional dimension of scalability. With its minimalistic design featuring only two check-valves as moving parts, the AGPM can be easily adapted as the board pressure generator for soft robots.

## Results and Discussion

The AGPM are fabricated as illustrated by **Figure 1**. The transducer material of the AGPM is an interconnected network of hollow graphene microtubes called aerographene. The aerographene is prepared as described elsewhere.[15, 16] Tetrapodal zinc oxide (ZnO) microparticles are molded into templates (20x20 mm with a thickness of 3-5 mm) and sintered (5 h at 1150 °C). The process creates highly porous (~94 %) networks of interconnected tetrapods. The templates are highly hydrophilic and can be infiltrated with an aqueous dispersion of 1.4 mg ml$^{-1}$ electrochemically exfoliated graphene (EG).[18]

Upon drying, the graphene deposits on the tetrapod arms. Repeating the infiltration 6 times results in a homogenous self-assembled thin film of graphene on the ZnO network. The samples are then wet-chemically etched to remove the underlying t-ZnO and then critical point dried. The resulting aerographene is a freestanding network of interconnected graphene microtubes with an ultra-low density of ~10 mg cm$^{-3}$ and an exceptionally high porosity (>99.99 %). **Figure 1a** shows a photograph of a finished aerographene sample with a cross-sectional SEM of the internal structure and schematics of the aerographene microstructure as additions. The hollow graphene microtubes are roughly 50 µm in length and 3 µm across with a wall thickness of about 25 nm.[15] The aerographene is then electrically contacted using conductive silver paste and brass rods and mounted into the pressurizer module. The module is sealed using a sheet of silicone rubber and a top lid. A schematic of the assembly with the individual parts and a photograph of an assembled AGPM are presented in **Figure 1b** and **Figure 1c**, respectively. With only two moving parts (the check-valves) that can be easily replaced, the AGPM is simple and cheap to maintain with minimum downtime. The minimalistic design can be easily miniaturized and adapted to fit the on-board electronics. The necessary electric power could be provided by an inductively chargeable battery. The AGPM would therefore be suitable as pressure supply in highly autonomous soft robots.

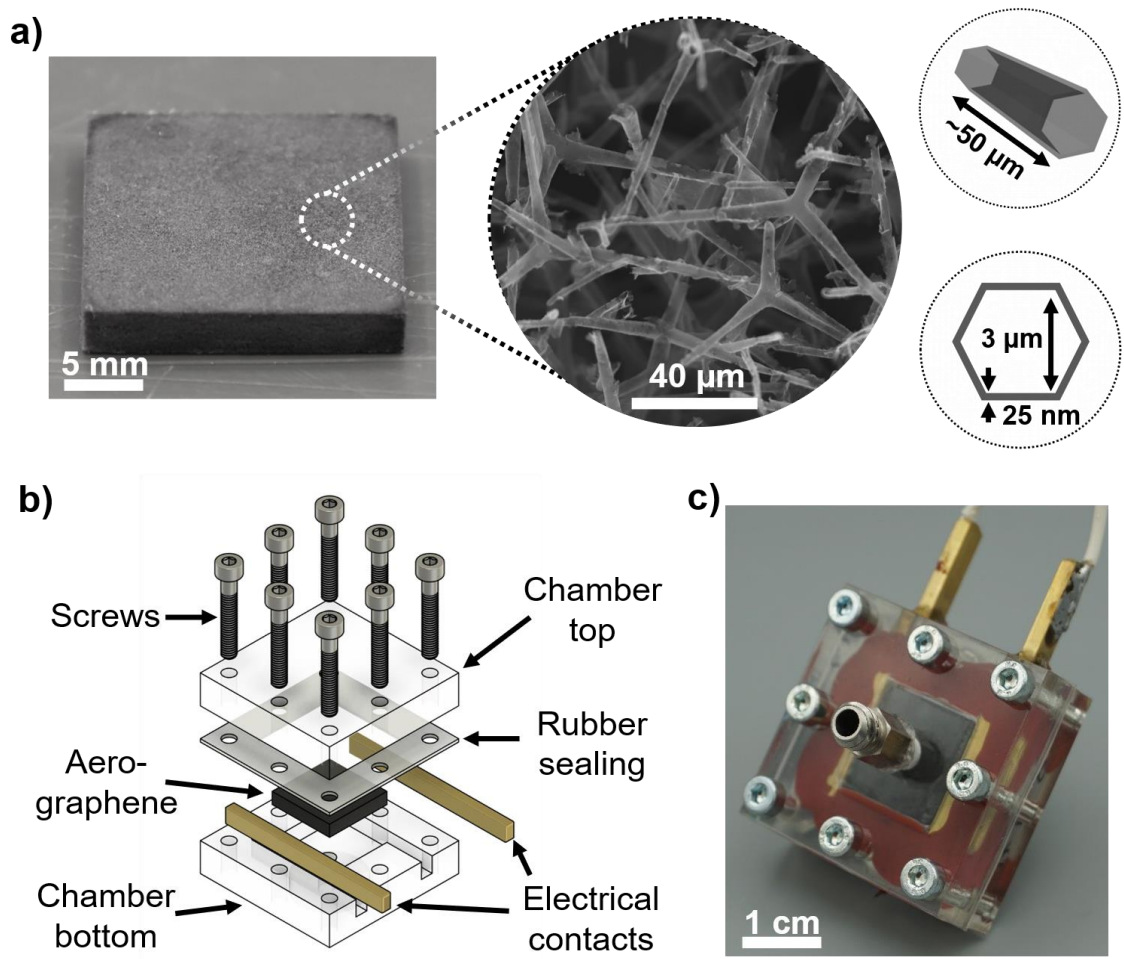

**Figure 1**: Fabrication of aerographene pressurizer module **a)** Photograph of aerographene sample with inserted SEM image and schematic of the freestanding hollow microtubes. **b)** Schematic of the aerographene pressurizer module. **c)** Photograph of assembled aerographene pressurizer module (AGPM).

Aerographene exhibits a low volumetric heat capacity (1-5 kJ m$^{-3}$ K$^{-1}$) and a high electric conductivity (~50 S m$^{-1}$).[16] In combination with the high thermal stability of graphene, these properties enable rapid Joule-heating at low power consumption. Under an oxidizing atmosphere, aerographene is stable up to temperatures of ~400 °C. Under inert gas, even higher temperatures of up to 800 °C are possible without considerable degradation.[19] **Figure 2** presents the results of a thermographic analysis of the Joule-heating behavior of the aerographene employed in the AGPM. All further reported values for electrical powers are averaged over the duration of the power pulse. **Figure 2a** shows a photograph of the sample as well as infrared images of 20x20x5 mm aerographene taken 30, 90, and 180 ms after the start of a heat pulse of 54 W for 100 ms. The material heats up rapidly and homogenously with only the boundary regions in contact to the brass rods being significantly colder than 300 °C. This is due to heat conduction to the contacts. The complete infrared video in slow-motion is supplied in **Video S1**. The histogram presented in **Figure 2b** was calculated from the temperature data of the IR image taken after 90 ms. The mean temperature of the sample was ~306 ± 59 °C, with ~70 % of the material showing temperatures above ~280 °C. A table with the temperature distribution data is provided in the supplemental information (**Table S5**). **Figure 2c** shows the maximum temperature of the 20x20x5 mm sample over time when power pulses of equal energy, but different pulse lengths are applied. The maximum temperature decreases with pulse length. While the AGPM reached a temperature of 438.67 ± 8.39 °C when ~5.2 J were applied over 50 ms, it only reached 353.67 ± 9.61 °C when the energy was applied over 200 ms. The effect is more pronounced for modules employing aerographene of higher thickness. In **Figure 2d** the temperature response of aerographene to power pulses adjusted to heat the material to 400 °C is shown. From fitting the cooling curves, it can be estimated that the aerographene cools down to 30 °C in less than 5 s (**Figure S3**). The energy required to heat the aerographene increases with pulse length. In contrary to the effect on the maximum temperature, the effect of the thickness on the energy required to heat the active material to 400 °C is smaller for thicker aerographene samples. A more detailed analysis of the power consumption and maximum temperature for varying pulse lengths is presented in the supplemental information (see **S2**). The long-term cyclability of a 20x20x5 mm aerographene sample is demonstrated in **Figure 2e**. The sample was repeatedly heated to ~400 °C by applying 50 W for 100 ms a total of 50 times. Between each pulse, the system was left to cool for 1400 ms. No significant changes in the temperature response of the aerographene were observed, indicating that no thermal degradation of the aerographene or the contact areas occurred.

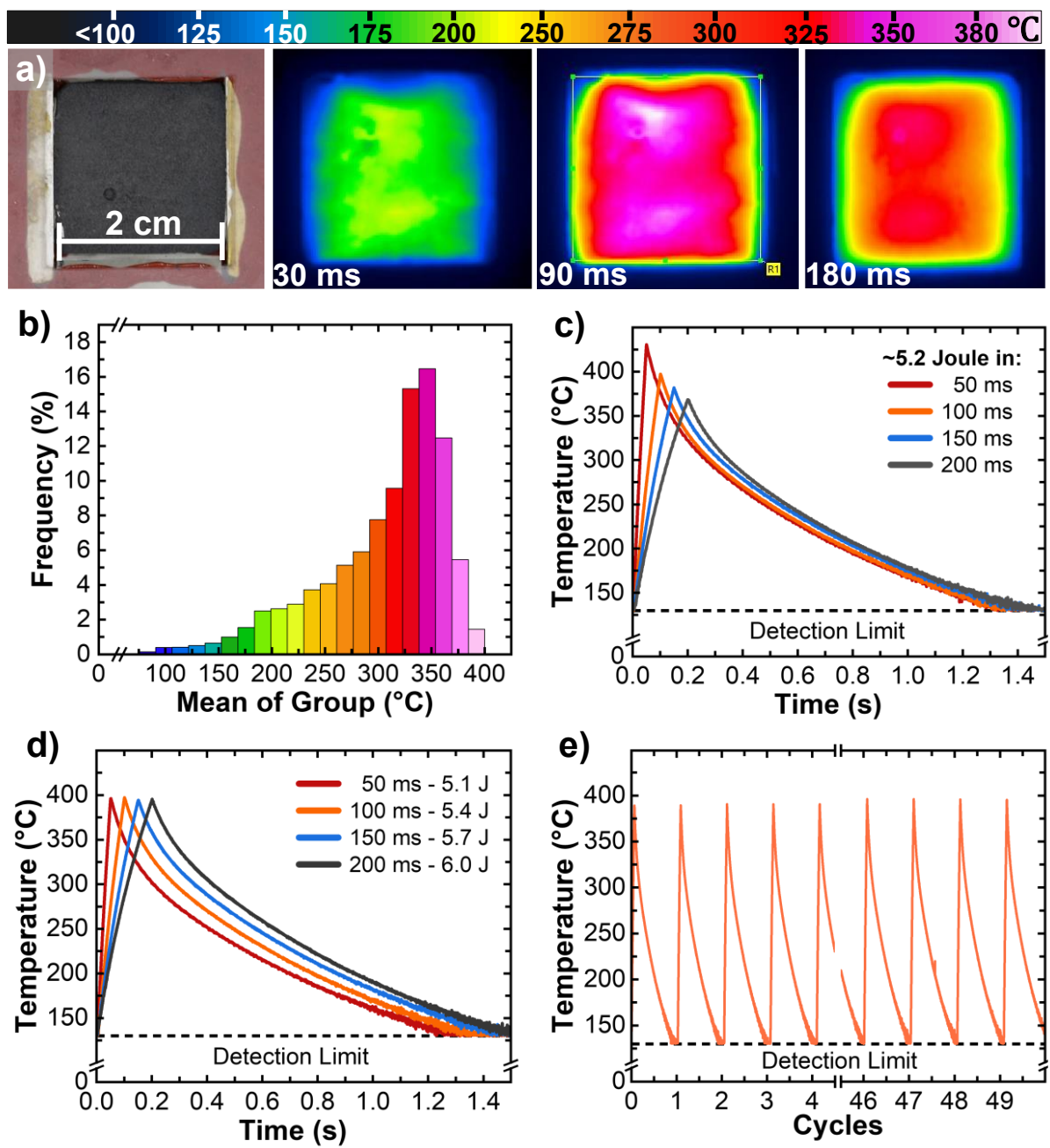

**Figure 2**: Joule-Heating of 20x20x5 mm aerographene sample. **a)** Infrared imaging during heating with 54 W – 12.6 V for 100 ms with subsequent cooling of 1400 ms. Images are taken after 30, 90 and 180 ms. **b)** Histogram for the infrared image taken after 90 ms. **c)** Time-temperature curves of aerographene when heated with pulses of ~5.2 J applied over varying pulse duration. **d)** Time-temperature curves of aerographene when heated to ~400 °C by applying power pulses of different length **e)** First and last four pulses of a total of 50 repeated cycles of heating with 50 W – 12.1 V for 100 ms with subsequent cooling of 1400 ms.

The open porous aerographene network has a high and fully accessible volumetric surface area (~0.14 m² cm⁻³). When the aerographene in the AGPM is Joule-heated, heat can be efficiently transferred to the air in and around the material. As the aerographene has a very small Knudsen number (≪1), the properties of the gaseous phase can be estimated to that of free gas.[16] The heated air can therefore move out of the material freely. These properties enable electrically powered repeatable air explosions (EPRAE) as demonstrated in previous works.[16] By outfitting the AGPM with check-valves on the in- and outlet, a unidirectional gas flow can be created. The AGPM can therefore be used as a pump.[16] When the AGPM is connected to a sealed container at the outlet, the system can be used for pressure generation. A detailed characterization of the pressure generation by a single AGPM is shown in **Figure 3**. The AGPM was outfitted with medical grade fluidic check-valves and connected to a pressure reservoir with a volume of ~4.2 cm³. A photograph of the setup used for the pressure generation is shown in **Figure 3a**. Joule-heating the aerographene transfers heat to the air in and around the material. Following the ideal gas law (**Equation 1**), the pressure in the AGPM increases proportional to the change in temperature. Due to the excess pressure in the module, the outlet check valve opens. The pressure in the module and the pressure reservoir equalizes by pushing air from the module into the reservoir. Once an equilibrium is achieved, the outlet check-valve of the module closes. After the heat pulse the pressure in the system decreases as the air cools down. Due to the increased amount of air in the reservoir, its pressure stays above the initial value. The pressure in the AGPM on the other hand falls below ambient pressure. Thereby, the inlet check-valve of the module opens, and cool air can flow in to reset the AGPM. The pressure generated by the AGPM is determined by the temperature increase of the air in and around the graphene caused by the Joule-heating. Repeating the procedure periodically causes a stepwise increase in pressure of the reservoir. **Figure 3b** presents the pressure evolution for periodically operated AGPM employing 20x20 mm of aerographene with varying thickness. The transducer material is operated with power pulses that, based on the results of the Joule-heating characterization (**Figure 2** and **S2**), will heat the aerographene to ~400 °C. The generated pressure varies for AGPM employing aerographene of different thicknesses. On the initial pulse, the 3 mm AGPM created 106 ± 1.9 mbar. The 4 mm and 5 mm AGPM generated significantly higher initial pressure of 138 ± 11.2 mbar and 132 ± 2.2 mbar, respectively. The pressure difference is smaller for each consecutive pulse until a maximum pressure is reached. After 10 pulses all modules reached their maximum pressure. The 3, 4, and 5 mm AGPM generated maximum pressures of 272 ± 15, 341 ± 30 and 316 ± 10 mbar, respectively. As there is a necessary gap between the aerographene and the housing (1 mm), the filling factor ($V_{aerographene}/V_{module}$) of the modules increases with the thickness of the employed transducer material. The calculated filling factors of the modules are listed in the Materials and Methods (**Table 1**) section below. Compared to the thicker modules, the 3 mm aerographene needs to heat more air relative to its volume. This could result in a lower temperature difference which would explain the lower pressure. Although the difference in filling factor would explain the lower maximum pressure for the 3 mm AGPM, it would be inconsistent with the results for the thicker modules. The 4 mm modules demonstrate a higher maximum pressure than the 5 mm modules even though there filling factor is lower. It is suspected that there is either an optimum filling factor between 48 and 56 %, or other factors like the surface to volume ratio of the active material are also relevant.

**Figure 3c** presents the pressure evolution for 4 mm AGPM modules that were operated with pulses of different lengths. The power of each pulse was set to heat the module to ~400 °C. (see **S2**). As already mentioned in the discussion of the Joule-heating behavior above, energy required to heat the aerographene to ~400 °C increases with the length of the power pulse. A pulse of 50 ms needed to be performed using 80 W and a pulse of 200 ms required 25 W of power. However, even though the longer pulse applies more energy, the maximum pressure reached is lower. The added energy is therefore not used to heat the air in the system but is likely lost through heat conduction to the electrical contacts and the housing of the module. In can be concluded, that the ratio between heat transferred to the air and heat transferred to the contacts decreases with pulse length. Shorter pulses are therefore favorable for pressure generation. As the pressure generated by the AGPM depends on temperature difference generated through the Joule-heating, the system needs to cool down in between pulses to reach maximum overpressure. **Figure 3d** shows the resulting pressure curves of 4 mm APGM pulsed 10 times with pulses of equal energy but different cooling intervals in between each pulse. When operated with 20 pulses per minute, the tested AGPM reached a maximum pressure of ~365 mbar in 30 s. At 30 and 40 pulses per minute, maximum pressures of ~350 mbar in 15 s and ~345 mbar in 10 s were achieved, respectively. The shorter the cooling interval, the higher the residual temperature of the aerographene at the start of the next pulse. Although higher starting temperature will result in a maximum temperature, the difference between starting and maximum temperature will be smaller. The maximum pressure reached by the AGPM therefore decreases with the cooling interval. However, the difference in maximum pressure smaller than one would expect from looking at the cooling behavior of the aerographene after heating (**Figure 2c-d**). This is most likely due to increased cooling from cool air flowing in after the pulse. Not just the module, but also the gas in the pressure reservoir needs time to normalize its temperature (see **S3**). The maximum reachable pressure at higher repetition rate could be further increased by enhancing the system's ability to cool. This could be achieved by, for example, using anodized aluminum for the casing of the AGPM. The high thermal conductivity of aluminum could improve the cooling rate of the module while the anodization would prevent short circuiting the aerographene. The aluminum would also provide the electromagnetic shielding necessary for on-board electronics in robotic systems. As demonstrated by **Figure 3e,** the pressure generation capability of the AGPM can be further improved by increasing the network density and thus the internal surface area of the active aerographene material. The pressure curves of AGPM fabricated with aerographene produced from templates with twice the t-ZnO density (0.6 g cm$^{-3}$) are compared to a reference produced with 0.3 g cm$^{-3}$. The modules with higher template density generated $373 \pm 28$ mbar, which is a 10 % higher maximum pressure than the reference. The modules reached the maximum pressure in the same number of pulses, indicating that the increased network density did not impact the volume of air heated. These results indicate that the aerographene itself offers additional potential for further optimization of the AGPM.

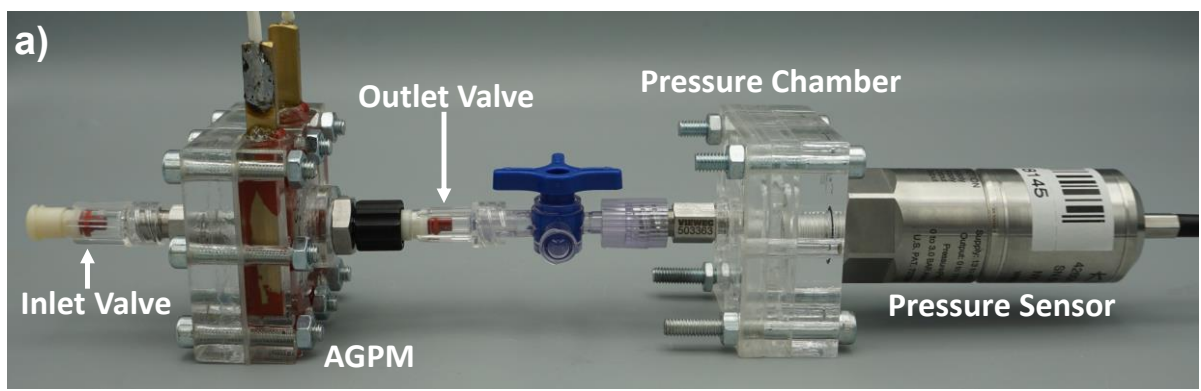
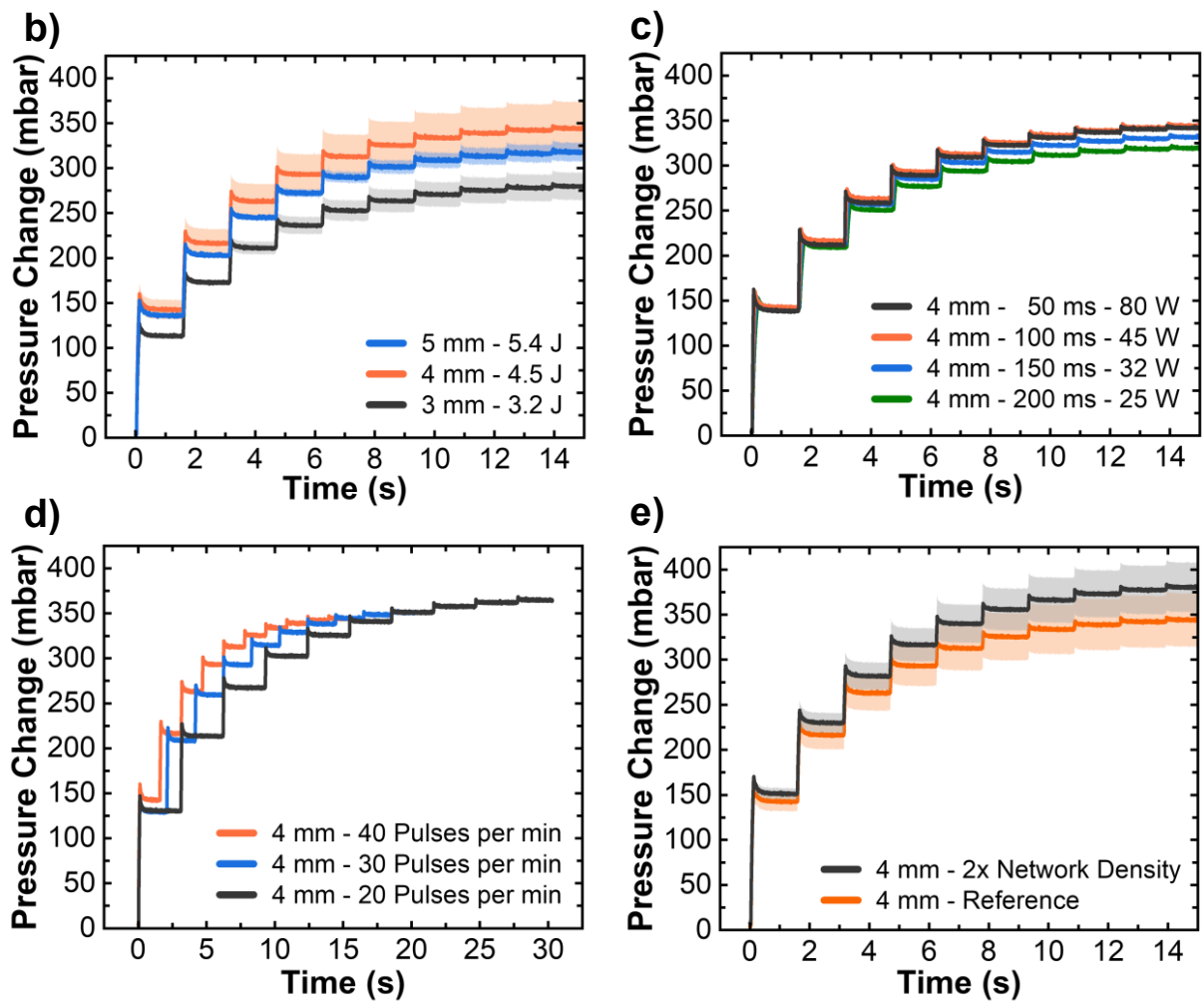

**Figure 3**: **a)** Photograph of pressure characterization setup. The AGPM is outfitted with medical grade check-valves and connected to a pressure reservoir featuring a high precision pressure gauge. **b)** Pressure curves for AGPM of different thickness heated to ~400 °C in 100 ms with subsequent cooling of 1400 ms. **c)** Pressure curves of 4 mm samples heated to ~400 °C with power pulses of various length. **d)** Pressure curves of 4 mm AGPM pulsed with 45 W for 100 ms at intervals of 20, 30 and 40 pulses per minute. **e)** Pressure curves for AGPM of two different network densities operated with 45 W for 100 ms at 40 pulses per minute.

The scalability of the AGPM concept is demonstrated in **Figure 4a** by connecting several modules together. APGM can be connected and operated parallelly or in series to achieve improved pressure results. By connecting multiple modules in series and pulsing them alternatingly (45 W for 100 ms at 40 ppm), the maximum pressure reached by the system is significantly increased. **Figure 4b** shows the pressure curves for 2 and 3 modules in series compared to a single module. In the 15 s a single module needs for maximum pressure, two modules in series generate 569 ± 34 mbar and three modules generate 624 ± 35 mbar of pressure. Based on exponential fitting performed on the pressure curves (see **S5**), the maximum reachable pressure is expected to be ~850 mbar for two and ~1450 mbar for three modules, respectively. Reaching the maximum pressure takes more time the more modules are in series. The time constant of the exponential fit increased from 3.15 s for a single module to 14 s and 28 s for two and three modules, respectively. A benchmarking of the AGPM concept was performed using three AGPMs with increased network density in series (see **S6**). The system reached a maximum pressure of almost 2000 mbar in about 2.5 min. The maximum pressure reached by two AGPM in series is more than doubled compared to a singular module. Assuming the percentual increase in pressure is constant in each consecutive module, the maximum pressure of two modules should not exceed ~800 mbar. However, this value is experimentally exceeded by 50 mbar, even though the first module is not pressurized to its maximum pressure between each pulse of the second module. This indicates that the transfer of energy from the aerographene to the gas phase increases with its pressure. Operating the thermopneumatic system on a pre-pressure could therefore also improve the performance. On the other hand, the higher the pressure, the more gas molecules are present. Therefore, the higher the pressure, the more energy is required to heat up the gas. As the supply power was kept constant for every module, this would explain why three modules in series show a lower maximum pressure than one would expect from this theory. When more and more units are operated in series, at some point the energy of the power pulse will become insufficient to significantly increase the temperature gas phase. To overcome this limitation, the supply power for higher order modules could be adjusted. Connecting multiple AGPM modules parallelly increases the pressure that can be generated per pulse. **Figure 4c** presents the pressure curves of two and three parallelly connected modules compared to a single system. Especially the pressure increase caused by the first pulse is significantly higher when multiple units are used parallelly. A more detailed analysis of the first 5 pulses is presented in the supplemental information (**S4**). While a single 4 mm module generates about 140 mbar on the first pulse, this value is increases to ~170 mbar for two and ~215 mbar for three modules. However, the maximum reachable pressure is not increased. As demonstrated by **Figure 4d**, both concepts can be combined to exploit their effects in tandem. Two parallel modules with another module in series show both, an increased pressure on the first pulses as well as an increased maximum pressure compared to a single module. By clever combination of multiple AGPMs in parallel and series a pressure generator can be created that is specifically adapted to its intended use case.

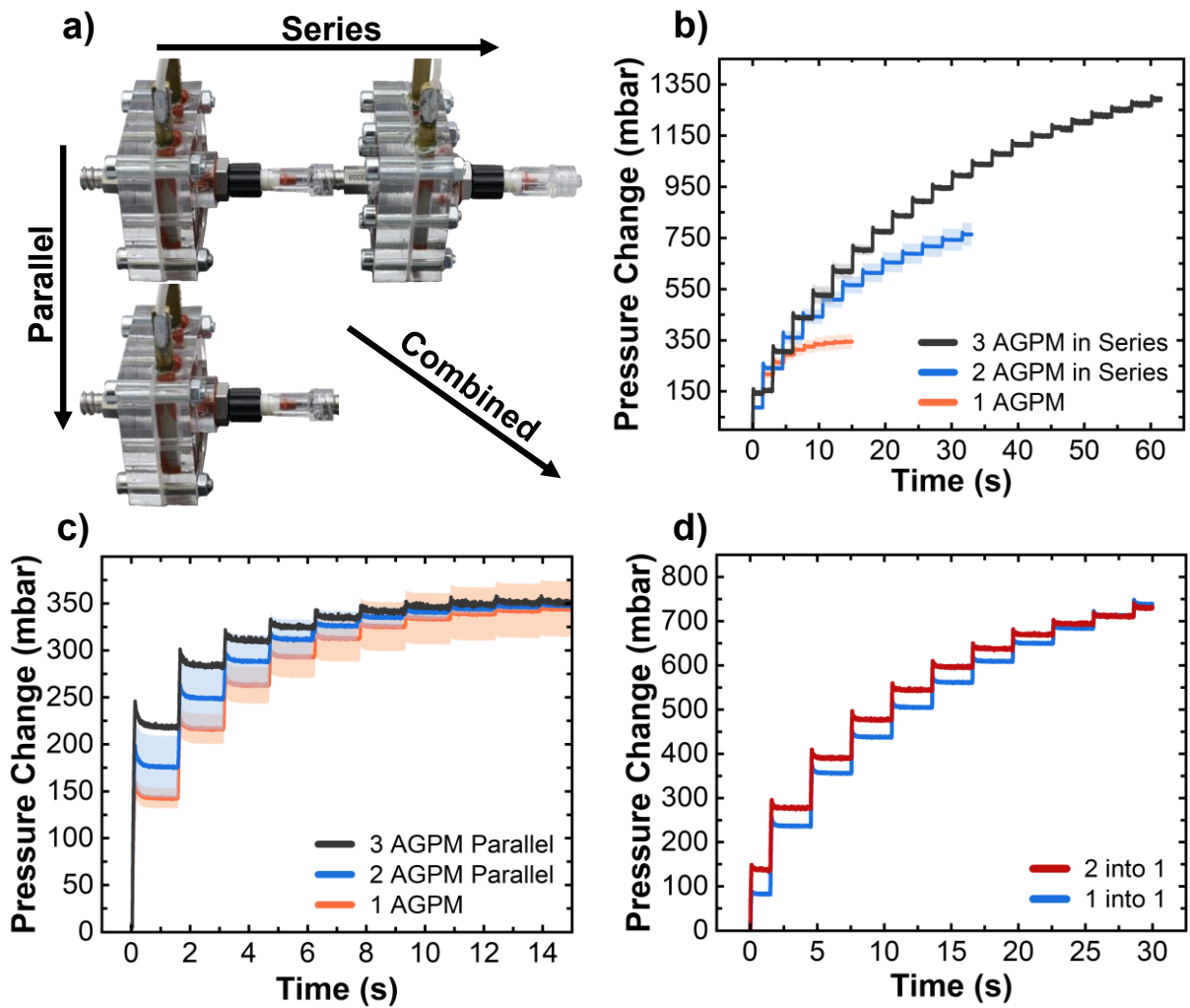

**Figure 4**: Scalability of the pressurizer concept. **a)** The pressurizer modules can be used in series, parallel, or a combination of both. **b)** Pressure generation of multiple modules in series. **c)** Pressure of multiple parallel modules. **d)** Pressure generation of a system consisting of two parallel modules with a third in series.

## Conclusion

In conclusion, this study presents a new type of thermopneumatic pressure generator using a hierarchical network of carbon microtubes called aerographene as transducer material. Compared to the pleated or coiled wires used as transducer material in previously reported thermopneumatic systems, aerographene has a significantly higher accessible surface area, lower heat capacity. Thereby, aerographene can be rapidly Joule-heated to 400 °C under air. Used as transducer material in the aerographene pressurizer module (AGPM), it enables thermopneumatic pressure generation with low response times (50-100 ms) and high repetition rates (~1 Hz). An AGPM employing 20x20x4 mm (1.6 cm$^3$) of aerographene is able to pressurize a reservoir of ~4.2 cm$^3$ to around 140 mbar with a single pulse of 80 W for 50 ms.

Through periodic operation of the AGPM at a frequency of 0.66 Hz, the pressure was further increased to approximately 360 mbar within just 10 seconds. It was further demonstrated, that several AGPM can combined either parallelly or in series for improved pressure generation. Operating AGPMs in parallel results in an increased pressure generated by the initial pulse. Three parallelly operated AGPM were capable of generating a pressure ~215 mbar in 100 ms. Connecting AGPMs in series and operating the alternatingly leads to an increase in the maximum pressure reached by the system. With three AGPM working in series, it was possible to pressurize 4.2 cm$^3$ of air to almost 625 mbar in 15 s or 2000 mbar in about 2.5 min. With most of the recently presented bioinspired pneumatic grippers operating on 5 mbar of pressure, having 4.2 cm$^3$ of air at ~2 bar is equivalent to almost 1700 cm$^3$ at 5 mbar. The AGPM should therefore be more than capable of providing the necessary pressurized air for a robot utilizing such grippers. Furthermore, the AGPM concept provides many opportunities for further optimization. For example, by using inert gas as the medium instead of air, the aerographene could potentially be heated to temperatures around 800 °C or even beyond, resulting in significantly higher pressures. Improving the AGPMs ability to cool after the heat pulse by, for example, using anodized aluminum for the casing of the module, could increase the possible repetition rates at high maximum pressures. With a plethora of possible ways to combine different AGPMs it is possible to create pressure generators specifically tailored for the designated use-cases. The minimalistic design of the AGPM with only two moving parts is easy to miniaturize and maintain. The AGPM could be easily adapted to circuit boards, making the concept an excellent fit as on-board pressure generator for autonomous soft robots.

## Materials and Methods

**Resource availability.** The tetrapodal zincoxide (t-ZnO) generated in this study will be made available in reasonable amounts on request to the corresponding author. There are restrictions to the availability of t-ZnO because of our limited production capability and our need to maintain the stock. We may require a payment and a completed materials transfer agreement if there is potential for commercial application. Large quantaties of t-ZnO are available for purchase from PhiStone AG (Mielkendorf, Germany). All 3D files and schematics necessary to reproduce the devices presented in this work will be made available by the corresponding authors upon request. All other materials are available from the respective sources quoted in this section.

**Fabrication of aerographene.** Zinc oxide tetrapods were produced using the flame transport synthesis approach described by Mishra, Kaps et. al..[20] A 2:1 mixture of Mowital B60H (Kuraray Europe GmBH, Germany) and zinc dust (Sigma-Aldrich Chemie GmBH, Germany) is placed in a furnace at 900 °C for 20 min. The resulting t-ZnO particles are harvested from the crucible manually. The t-ZnO particles are compressed into 20x20 mm templates of various thickness with a densities of 0.3 g cm$^{-3}$. Additional templates with a network density of 0.6 g cm$^{-3}$ are prepared in 20x20x4 mm. The templates are sintered at 1150 °C for 5 h to increase mechanical stability. The templates are then infiltrated with an aqueous dispersion of electrochemically exfoliated graphene. The dispersion is prepared similar as described elsewhere[18] and kindly provided by Sixonia Tech GmbH (Dresden, Germany).

The dispersion is diluted to 1.4 mg ml$^{-1}$ with deionized water before being tip sonicated with 3 s pulses for a total of 10 min and 13.16 kJ using a Sonoplus HD4100 (Bandelin, Germany). The dispersion is drop infiltrated into the templates until their free volume is filled. After infiltration, the samples are dried for 4 h at 50 °C. The procedure is repeated a total of 7 times. The processed templates are then placed in 10 % hydrochloric acid for 24 h to etch out the ZnO. The resulting aeromaterial samples are then washed 3 times each in deionized water and pure ethanol before being dried using EM CPD300 critical point dryer (Leica, Germany). The aerographene is then electrically conducted using brass rods and conductive silver paste (Acheson 1415, Plano, Germany) before being mounted into CNC machined sample holders. The base of the aerographene pressurizer module is sealed using highly temperature resistant silicone (Pattex Ofen und Kamin Spezial Silikon, Henkel, Germany). A silicone rubber sealing and a screwed on top lid completes the module.

**Table 1**: Cavity and respective sample volume with the corresponding filling factor for AGPM employing aerographene of varying thickness.

|  | 3 mm | 4 mm | 5 mm |
|---|---|---|---|
| **Cavity Volume (cm³)** | 2.52 | 3.02 | 3.47 |
| **Sample Volume (cm³)** | 1.20 | 1.60 | 2.00 |
| **Filling Factor (%)** | 47.62 | 53.0 | 57.6 |

**Setup for rapid Joule-heating.** Joule-heating the aerographene is performed using a self-designed rapid heating setup. For temperature measurements, a high-speed Metis H318 pyrometer (SensorTherm GmbH, Germany) is used. An EA PS9080-60 is used as power supply. Data recording and setup control are performed by an USB2537 High-Speed DAQ Board (Measurement Computing Corporation, USA) in combination with a self-programmed Python software.

**Characterization of Joule-heating behavior.** AGPM with 20x20 mm aerographene of varying thickness are Joule-heated using the setup described above. The top-lid is removed and the aerographene is covered by a codial window with high infrared permeability (SFK16, VAb Vakuum-Anlagenbau, Germany). The samples are heated to ~400 °C by applying short power pulses of 50-200 ms. The power required for each pulse length is recorded. Each sample is also heated with the energy required to heat the sample to 400 °C in 100 ms but applied over 50, 150 and 200 ms. The maximum temperature reached by each pulse is recorded. Additional experiments are performed on 4 mm AGPM fabricated using aerographene with twice the t-ZnO template density (0.6 g cm$^{-3}$). The experimental parameters are the same as for the 4 mm samples with standard density.

**Thermographic Analysis.** An unsealed aerographene pressurizer module with a sample of 5 mm thickness is Joule-heated with 54 W – 12.6 V for 100 ms using the rapid heating setup described above. Infrared images and additional temperature data are recorded using a HD VarioCAM infrared camera (InfraTec, Germany). Evaluation of the camera data is performed using the thermography software IRBIS 3.1 (Infratec, Germany). A histogram of the camera temperature data is generated using a self-written Python program.

**Investigation on cycling stability.** To show cyclic stability of the aeromaterial, a 20x20x5 mm sample is primed by Joule-heating it to ~400 °C a total 5 times. The sample is then left to cool for 2 min. Afterwards, the sample is Joule-heated to ~400 °C by repeatedly applying ~50 W – 12.1 V for 100 ms with a subsequent cooling period of 1400 ms. A total of 50 cycles is performed.

**Pressure generation.** Each aerographene module is connected to two medical check-valves (Infuvalve - Braun, Germany) to form a pressurizer module. The pressurizer module is then connected to a pressure reservoir with a volume of 4.3 ml using a medical 3-way luerlock valve (Braun, Germany).

The pressure reservoir is outfitted with a Type 4260A pressure sensor (Kistler Instrument Corporation, USA). The pressure generated through Joule-heating pressurizer modules with samples of a t-ZnO template density of 0.3 g cm$^{-3}$ and varying thickness is measured. Based on the results of the characterization of the Joule-heating behavior (**S1**), the applied electrical powers for each respective pulse length listed in **Table 2** are chosen. The samples are pulsed a total of 10 times every 1.5 s. The pressure in the pressure reservoir is measured continuously. The experiments are also performed for 4 mm pressurizer modules of twice the ZnO template density (0.6 g cm$^{-3}$) with the same parameters as the standard 4 mm modules.

**Table 2:** Power used for pulses of varying length to Joule-heat pressurizer modules with 20x20 mm AGPM of different thickness.

|         | 3 mm  | 4 mm  | 5 mm  |
|---------|-------|-------|-------|
| **50 ms**  | 60 W  | 80 W  | 100 W |
| **100 ms** | 34 W  | 45 W  | 54 W  |
| **150 ms** | 25 W  | 32 W  | 37 W  |
| **200 ms** | 20 W  | 25 W  | 30 W  |

The influence of the cooling interval between each pulse is determined by pulsing the pressurizer modules with different frequencies. 4 mm pressurizer modules are pulsed 10 times with 45 W for 100 ms. The modules are pulsed every 1 s, 1.5 s, and 2 s, respectively.

**Parallel operation of pressurizer modules**. AGPM of 4 mm thickness are combined via 3-way luerlock valves before being connected to the pressure reservoir as described above. The modules are operated simultaneously at 45 W for 100 ms with cooling intervals of 1400 ms. The modules are pulsed a total of 10 times. The pressure in the pressure reservoir is continuously recorded. Parallel operation of two and three AGPM is analyzed, respectively.

**Serial operation of pressurizer modules.** Two 4 mm AGPM are connected in series and attached to the pressure reservoir. The modules are operated alternatingly in several ratios, pumping the first module 1-5 times before pumping the second module once. Power pulses of 45 W for 100 ms are applied. The cycle is repeated several times while the pressure in the reservoir is continuously recorded.

Three modules in series are operated in a 1_0_1 to 0_1_0 ratio, meaning the first and last module are pulsed simultaneously once, before the middle module is pulsed once. Power pulses of 45 W for 100 ms are applied. The pressure in the reservoir is recorded continuously over several cycles.

**Combined operation of pressurizer modules.** Two parallel 4 mm AGPM are connected in series with a third 4 mm AGPM and the pressure reservoir. The parallel modules are operated simultaneously once before the module in series is pulsed. All modules are pulsed with 45 W for 100 ms with subsequent cooling of 1400 ms. The cycle is repeated several times while the pressure is recorded continuously.

**Electron Microscopy.** Scanning electron microscopy on several samples is performed using a Zeiss Supra 55VP with an in-lens detector.

## Acknowledgement

We acknowledge funding by the Deutsche Forschungsgemeinschaft (DFG) under contracts AD 183/12. This project has received funding from the European Union's Horizon 2020 Research and Innovation Program under grant agreement No GrapheneCore3 881603. We would also like to thank Dr. Martin Lohe from Sixonia Tech GmbH for providing the exfoliated graphene dispersion used in this work.

## Author Contributions

The AGPM concept was envisioned by J.B., R.A., A.R., J.R. and F.S.. R.A and F.S. secured the necessary funding. The required resources were provided by R.A.. The undertaking was supervised by F.S., R.A., as well as S.K.. Project administration was handled by A.R., J.R. and F.S.. The methodology for this work was developed by A.R., E.G., F.S., J.R. and M.M.. Any custom software was programmed by E.G., S.K. and A.R.. The investigations were carried out by A.R., and J.R.. The acquired data was curated by A.R., E.G and F.S. before being validated by A.R., J.R., E.G. and F.S.. Formal analysis was performed by A.R., J.R., E.G., F.S., and R.A.. The results were visualized by A.R., J.R., E.G., F.S. and M.M.. The first draft was devised by A.R. in cooperation with F.S.. All of the authors contributed to reviewing and editing the manuscript.

## Declaration of Conflicting Interests

The authors declare that there is no conflict of interests.

# Supporting Information

## S1 – Fabrication of Aerographene

**Figure S1** presents more details on the aerographene production. **Figure S1a** shows a photograph of a20x20x3 mm t-ZnO template after molding and sintering. The internal structure with the interconnected network of tetrapods is shown in the cross-sectional SEM image shown in **Figure S1d.** The template is infiltrated with an aqueous dispersion of graphene. A photograph of a partially filled template undergoing the infiltration process is presented in **Figure S1b**. The infiltration process is repeated 7 times to achieve a homogenous graphene thin film. A cross-sectional SEM image of the coated tetrapod network is presented in **Figure S1.** The sample is then etched in HCl overnight before being washed and dried using a critical point dryer. **Figure S1c** shows a photograph of a finished aerographene sample. A close-up SEM image of the hollow graphene microtubes can be seen in **Figure S1f**.

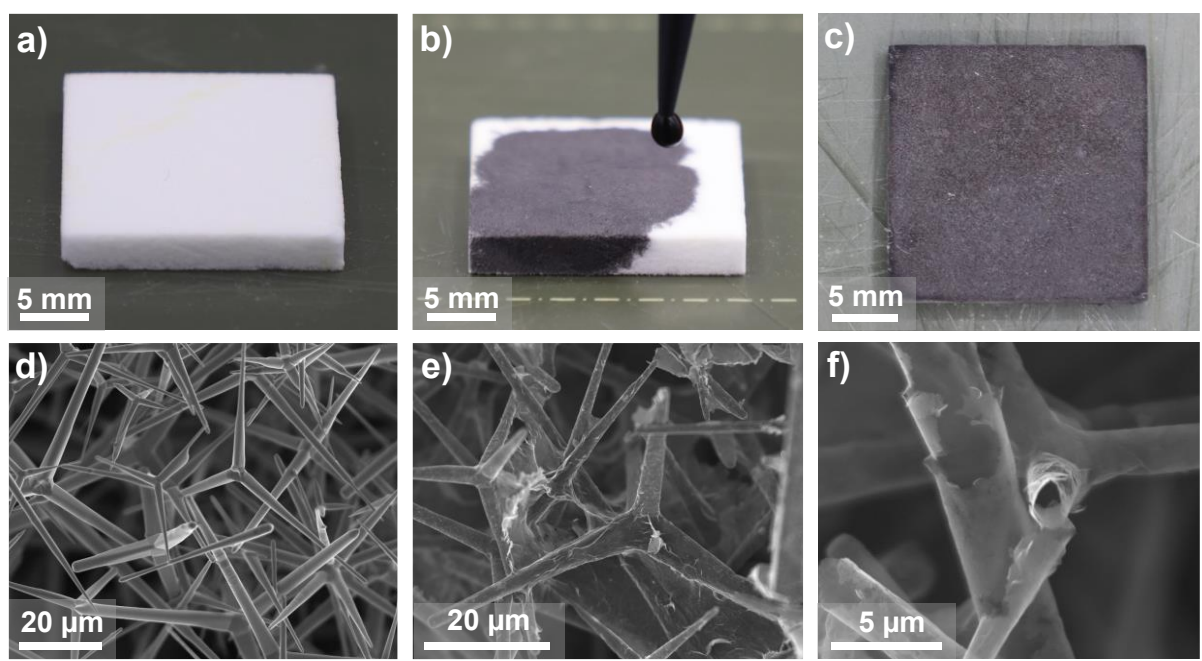

**Figure S1:** Preparation of 20x20x3 mm aerographene sample **a)** Template network of tetrapodal ZnO with a density of 0.3 g cm$^{-3}$ **b)** Partially filled ZnO template during drop infiltration process. **c)** Photograph of finished aerographene sample. Cross-sectional SEM images of **d)** ZnO template with a density of 0.3 g cm$^{-3}$, **e)** graphene coated ZnO network and, **f)** freestanding aerographene microtubes.

# S2 – Additional Joule-heating data

**Table S1** shows the results of the characterization of the Joule-heating behavior of pressurizer modules with varying thickness. The power required to heat the samples to 400 °C with pulses of different length were recorded.

**Table S1:** Electrical power required to heat aerographene samples to ~400 °C for varying sample thickness and pulse length.

|        | 3 mm          | 4 mm          | 5 mm            |
|--------|---------------|---------------|-----------------|
| 50 ms  | 61.5 ± 9.2 W  | 80.1 ± 3.2 W  | 110.64 ± 18.84 W |
| 100 ms | 33.9 ± 4.0 W  | 45.5 ± 1.2 W  | 60.11 ± 10.90 W |
| 150 ms | 24.6 ± 2.4 W  | 32.3 ± 1.0 W  | 42.92 ± 8.78 W  |
| 200 ms | 19.5 ± 1.7 W  | 25.7 ± 0.2 W  | 33.79 ± 6.43 W  |

**Figure S2a** shows the energy required to heat samples of varying thickness to 400 °C versus the length of the applied power pulse. When the energy is normalized to the sample volume, the values shown in **Figure S2b** is received. The energy required to heat the sample to ~400 °C increases with pulse length. The effect is less pronounced for samples of higher thickness.

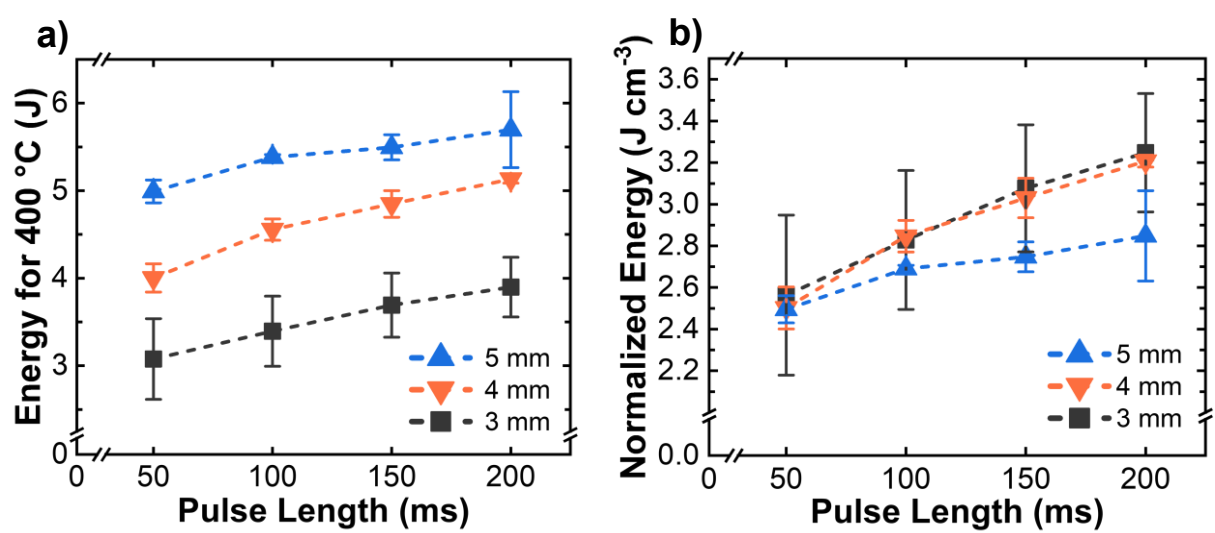

**Figure S2:** Results of the characterization Joule-heating behavior of the pressurizer modules. **a)** Energy required to heat the sample to ~400 °C versus length of power pulse. **b)** Energy data normalized to the sample volume.

Taking the energy required to heat each sample to ~400 °C in 100 ms as reference, the maximum temperature reached by pulses of equal energy, but different length was measured. The results of these experiments are presented in **Table S2.** The maximum temperature of a pulse with given energy decreases with pulse length.

**Table S2:** Maximum temperature reached when the energy required to heat the samples to ~400 °C in 100 ms is applied in different pulse lengths.

|        | 3 mm              | 4 mm              | 5 mm              |
|--------|-------------------|-------------------|-------------------|
| **50 ms**  | 433.71 ± 10.34 °C | 429.67 ± 8.96 °C  | 438.67 ± 8.39 °C  |
| **100 ms** | 399.42 ± 4.31 °C  | 397.17 ± 2.40 °C  | 396.67 ± 3.06 °C  |
| **150 ms** | 381.85 ± 11.73 °C | 386.67 ± 6.92 °C  | 370.33 ± 4.51 °C  |
| **200 ms** | 364.42 ± 13.51 °C | 374.83 ± 10.20 °C | 353.67 ± 9.61 °C  |

**Table S3** shows the Joule-heating characteristics of 4 mm pressurizer module with twice the t-ZnO template density compared to standard template density. The mean values of the required power to heat the aerographene to 400 °C are slightly higher for the modules with increased template density. However, as their standard deviations are also significantly larger, it is not admissible to make reliable statements here. The maximum temperature reached by pulses of normalized energy applied over different pulse lengths decreases with pulse length. The effect is more pronounced than for the 4 mm modules with standard template density.

**Table S3:** Results of characterizing the Joule-heating behavior of 4 mm samples with twice the t-ZnO template density.

|         | **Power (W)**     | **Temperature**   |
|---------|-------------------|-------------------|
| **50 ms**  | 88.27 ± 17.83 W   | 438.67 ± 6.11 °C  |
| **100 ms** | 48.86 ± 7.62 W    | 397.67 ± 2.31 °C  |
| **150 ms** | 36.06 ± 4.95 W    | 374.33 ± 6.92 °C  |
| **200 ms** | 28.53 ± 3.56 W    | 359.67 ± 8.38 °C  |

**Table S4**: Temperature data extracted from IR image taken after 90 ms of applying 54 W electrical power to a 20x20x5 mm aerographene sample.

| Temperature (°C) | Counts | Percentage (%) |
| --- | --- | --- |
| 68 | 6 | 0.03 |
| 83 | 25 | 0.14 |
| 99 | 71 | 0.39 |
| 114 | 75 | 0.41 |
| 130 | 93 | 0.51 |
| 145 | 118 | 0.64 |
| 160 | 184 | 1.00 |
| 176 | 284 | 1.55 |
| 191 | 458 | 2.50 |
| 207 | 481 | 2.62 |
| 222 | 531 | 2.89 |
| 238 | 684 | 3.73 |
| 253 | 747 | 4.07 |
| 269 | 943 | 5.14 |
| 284 | 1085 | 5.91 |
| 300 | 1424 | 7.76 |
| 315 | 1757 | 9.57 |
| 330 | 2811 | 15.31 |
| 346 | 3023 | 16.47 |
| 361 | 2291 | 12.48 |
| 377 | 1003 | 5.46 |
| 392 | 266 | 1.45 |

To receive metrics on the cooling behavior of aerographene after being Joule-heated, an exponential fit was applied to a mean of the temperature curves presented in Figure 2d of the main manuscript. The resulting fit and its function is presented in **Figure S3**.

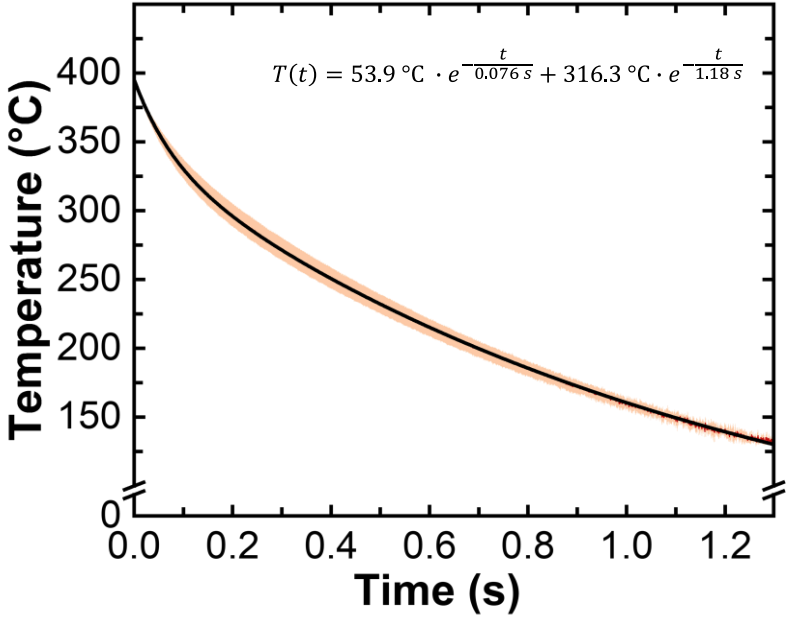

**Figure S3**: Exponential fit for cooling of 20x20x5 mm of aerographene after Joule-heating to 400 °C

**Figure S4** shows a plot of the conductivity of the aerographene versus its temperature. As graphene has a negative thermal coefficient of resistance, the conductivity of the material increases by 0.025 S m$^{-1}$ K$^{-1}$.

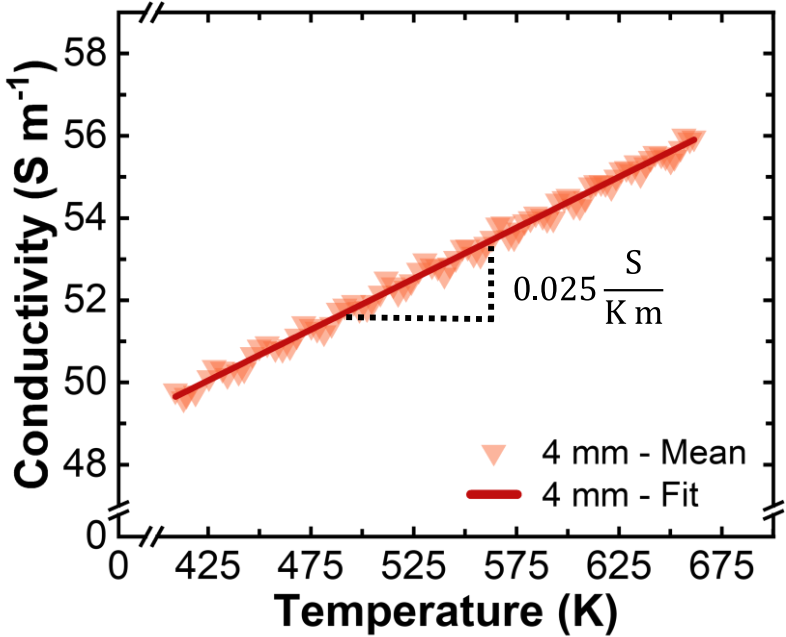

**Figure S4:** Conductivity of aerographene samples versus temperature and linear fit.

## S3 – Intercooling

As illustrated by **Figure 3d** of the manuscript, the aerographene pressurizer module needs time in between the heat pulses to reach the maximum possible overpressure. The time is necessary for both, the aerographene and the air in the final pressure reservoir to cool down. **Figure S5** shows the pressure generated by a 4 mm pressurizer module with an insert showing a zoomed-in plot of the first heat pulse. After each pulse, the pressure in the pressure reservoir decreases exponentially over a time of ~1 s with a time constant of 0.15 s.

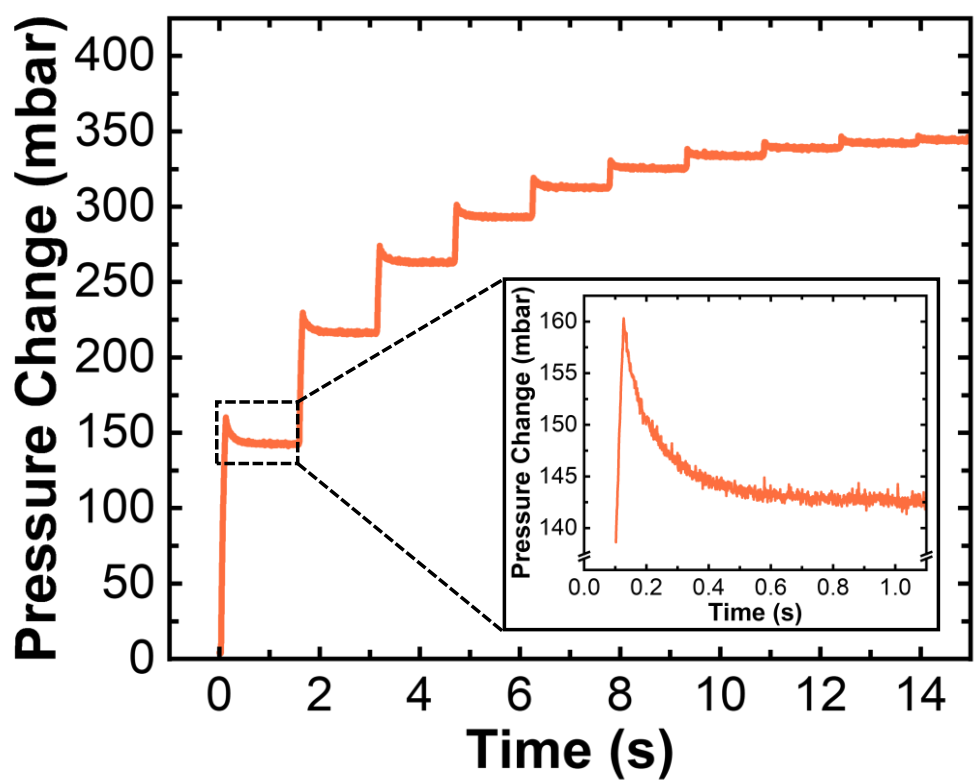

**Figure S5:** Pressure curve of 4 mm aerographene pressurizer module heated with pulses of 45 W for 100 ms with subsequent cooling of 1400 ms. The insert shows the pressure decrease due to cooling of the air in between peaks.

## S4 – Parallel AGPM

When samples are operated parallelly, the pressure generated by the initial pulse increases significantly. Although following a linear relationship, the pressure increase is not directly proportional to the number of modules used. Furthermore, the maximum pressure reached by the multiple units in parallel is not increased compared to a single module. The increase in pressure caused by the pulses following the first one, does therefore decrease with the number of units used. **Figure S6a** presents a close up on the 5 first pressure pulses of periodically operating multiple AGPM in parallel. **Figure S6b** shows the pressure increase per pulse as a bar graph for a more direct comparison.

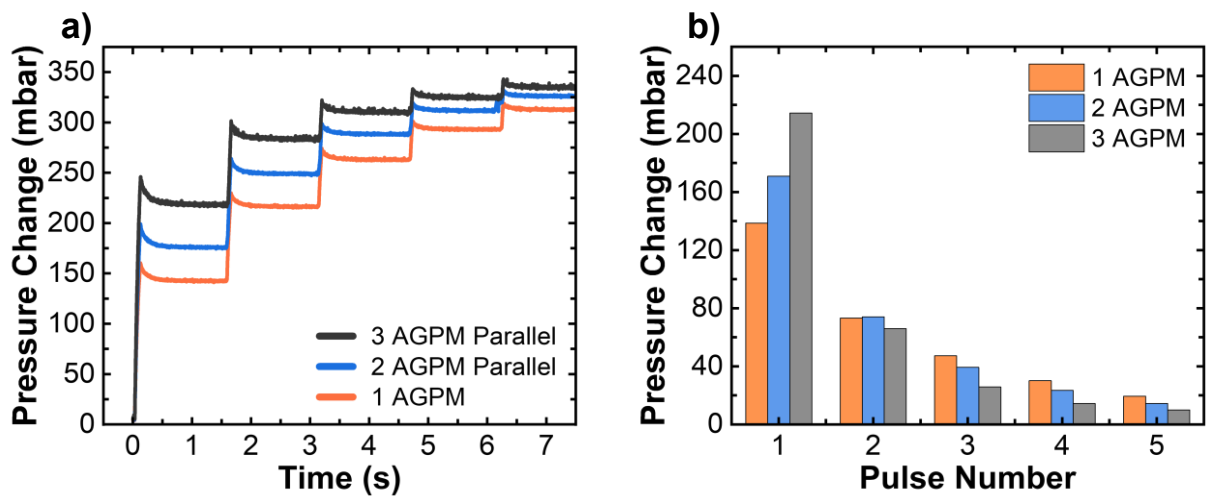

**Figure S6:** Pressure change per pulse for parallel operated pressurizer modules. **a)** Pressure versus time for the first 5 pulses. **b)** Pressure generated by each of the first 5 pulses of up to three parallel modules compared.

## S5 – AGPM in Series

To make predictions on the scalability of the AGPM in series, the pressure curves were fitted using the exponential association function in **Equation S1**.

$$P = P_0 + A_1\left(1 - e^{\frac{-t}{\tau}}\right) \quad (S1)$$

For the fit, the pressure after the cooling interval between pulses is used. **Figure S7a** shows the measured pressured data for a 4 mm pressurizer module operated with 45 W for 100 ms. The data points used for fitting are marked and the resulting fit curve is added. **Figure S7b** presents the fitting results for 4 mm pressurizer modules operated in series.

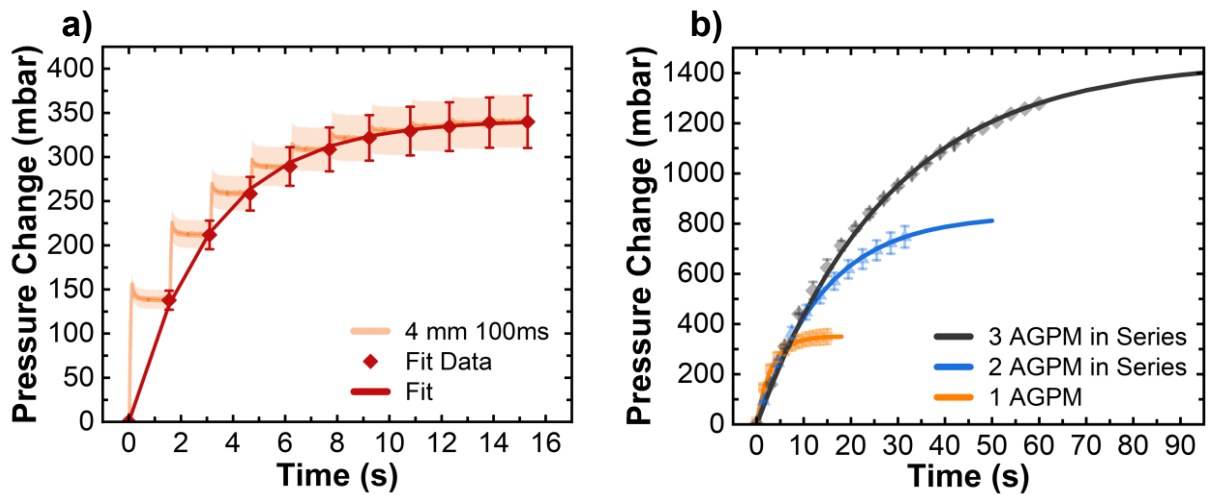

**Figure S7:** Fitting of pressure curves to predict maximum overpressure. **a)** Model validation on a single 4 mm sample pulsed with 45 W for 100 ms. **b)** Fitting of the pressure curves of one, two and three 4 mm samples in series. The two samples were pulsed in 1_0 to 0_1 scheme with 45 W for 100 ms. The three samples were operated in a 1_0_1 to 0_1_0 at the same pulse parameters.

From the exponential fits in **Figure S7** the time constants were extracted. These time constants are listed in **Table S5** below.

**Table S5:** Time constants of exponential fits performed pressure curves of multiple AGPM in series.

| Units in Series | A (mbar) | τ (s) |
| --- | --- | --- |
| 1 | 342 | 3.15 |
| 2 | 835 | 14 |
| 3 | 1450 | 28 |

## S6 – Benchmarking

Based on the results of the pressure characterization of the singular pressurizer modules, the most promising configuration was selected to benchmark the maximum pressure of the pressurizer system. Three 4 mm pressurizer modules with increased network density were operated in series with pulses of 45 W for 100 ms every 1.5 s. The units were pulsed in a 1_0_1 to 0_1_0 scheme until the pressure is no longer significantly increased. **Figure S8** shows the resulting pressure curve.

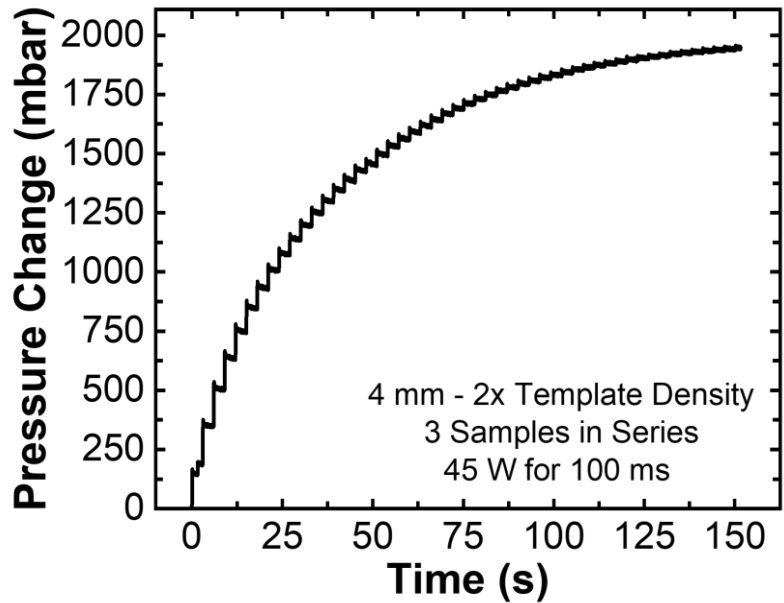

**Figure S8:** Benchmarking of EPRAE pressurizer system. Three 4 mm with increased template density were operated in series with 45 W for 100 ms every 1.5 s. The samples were operated in a 1_0_1 to 0_1_0 scheme.